\documentclass{amsart}
\usepackage{amssymb,stmaryrd}
\usepackage{amsfonts}
\usepackage{amstext}
\usepackage{algorithmic}
\usepackage{algorithm}
\usepackage{graphicx}
\usepackage{epstopdf}
\usepackage[all]{xy}
\usepackage{MnSymbol}

\newcommand{\Z}{\Bbb Z}\newcommand{\R}{\Bbb R}\newcommand{\C}{\Bbb C}
\newcommand{\extd}{{\rm d}}
\newcommand{\del}{\partial}
\newcommand{\id}{{\rm id}}
\newcommand{\tens}{\otimes}
\newcommand{\eps}{\epsilon}
\newcommand{\CC}{{\mathcal{C}}}

\newcommand{\CR}{{\mathcal{R}}}

\newcommand{\CV}{{\mathcal{V}}}
\newcommand{\ra}{{\triangleleft}}
\def\bicross{{\blacktriangleright\kern-2.9pt\triangleleft}}
\def\cobicross{{\triangleright\kern-2.9pt\blacktriangleleft}}
\newcommand{\cg}{{\mathfrak g}}
\newcommand{\cm}{{\mathfrak m}}
\newcommand{\cx}{{\mathfrak x}}

\newtheorem{example}{Example}

\begin{document}

\title{On the emergence of the structure of Physics}

\author{
S. Majid}

\address{School of Mathematical Sciences\\
Queen Mary University of London\\ 327 Mile End Rd, London E1 4NS, UK}

\thanks{Based on my talk at the {\em Workshop on Hilbert's 6th Problem}, Leicester, May 2016. }


\email{s.majid@qmul.ac.uk}

\begin{abstract}
We consider Hilbert's problem of the axioms of Physics at a qualitative or conceptual level. This issue is more pressing than ever as we seek to understand how both General Relativity and quantum theory could emerge from some deeper theory of quantum gravity,  and in this regard  I have previously proposed a {\em principle of self-duality} or {\em quantum Born reciprocity} as a key structure. Here I outline some of my recent work around the idea of quantum spacetime as motivated by this non-standard philosophy, including a new toy model of gravity on a spacetime consisting of four points forming a square. 
\end{abstract}
\maketitle 


\section{Introduction}\label{sec1}

Relative realism is a general philosophy that reality is like pure mathematics, created by decisions to work within certain axioms or assumptions \cite{Ma:pri,Ma:spbook,Ma:qg1}. To the extent that we are not conscious of this, to that extent we experience the reality created by those axioms. To the extent that we {\em are} aware, we transcend that level of `reality' but the fact that those axioms were possible, together with all the substructure they contain, is an element of a larger reality in which that was just one path we could have taken. This makes reality relative to your point of view, which is not necessarily a bad thing given the Copenhagen interpretation of quantum mechanics.

In this point of view, Physical Reality as we know it should be characterised or in some sense created by the decision to adopt certain axioms or assumptions. The difference with most mathematical subjects is that we don't {\em apriori} know what the axioms are but are working backwards to find them. My thesis in \cite{Ma:pri,Ma:spbook,Ma:qg1} was that if we eventually succeed then we will in fact uncover a characterisation of what it is to be a physicist. And knowing this, one can anticipate that one of the central axioms of Physics should turn out to be rooted in the scientific method, which one can formulate as a dual relationship between theory and experiment.

It is not clear to what extent Hilbert himself would have agreed with the above. Clearly, in asking for  `axioms of physics' in his 6th problem, Hilbert took the view as we do that there could indeed be axioms, which I would see as more than just a reductionist view that many modern-day theoretical physicists take in any case that physics can be subject to fundamental principles and ultimately boiled down to a single unified theory guided by those principles.  As explained in an excellent modern account \cite{Cor}, it is likely that Hilbert's ambition in the 6th problem was to elevate such axioms to a casting role comparable to the complete and independent set of axioms that he eventually found in \cite{Hil:gru} for geometry. What we do with the assumption that axioms exist, however, is very different. In relative realism we see such axioms as casting physics as a branch of pure mathematics with Physical Reality `created' by our adopting them, while Hilbert's position was the exact opposite that geometry was an empirical subject more like physics with its reality already out there as discovered by experience, i.e. not that physics was more like geometry as pure mathematics but that geometry was more like physics as empirical.  Hilbert saw the role of axioms for physics as a way to put it on a firmer footing, or in his words as quoted in \cite{Cor}, to support and fortify `{\em when signs appear that the loose foundations are not able to sustain the expansion of the rooms}'. The metaphor here was that Science is a growing edifice of different rooms in which the different branches operate. The physics here is already out there, we approach it patchily and then mathematics has a retrospective clarifying role, although also an explanatory role if, ideally, one finds axioms that lead to exactly the physics that is observed. 

In fact, Hilbert took the view that mathematics in general is `{\em not like a game whose tasks are determined by arbitrarily stipulated rules}'\cite{Hil, Cor} but rather that good axioms are part of a pre-existing structure of `mathematical reality' as I would put it.  In my experience, many mathematicians would also agree with this, although many would not. For example, the French mathematician Dieudonn\'e  famously argued\cite{Die} the other side that mathematics was more like a game of chess in which there is no absolute truth since the rules are abritrary. In relative realism we do take the  `mathematics is out there' view but we {\em also} take the Dieudonn\'e view except that we do not see it as perjorative. Thus,  there is a reality experienced by chess-players as they experience the restrictions of chess while playing, in some sense `created' or carved out by those restrictions, but at the same time the rules of chess are not arbitrary and constitute an `empirical fact' although at a higher level as an element of the reality of possible board games as experienced by designers of board games\cite{Ma:pri}. This gives a hierarchical `room within rooms' structure to our experience of reality a little different from Hilbert's analogy. Exiting a room by dropping an axiom takes us into a bigger room in which the door we just came out of is just one of the points of interest \cite{Ma:hel}. 

In summary, Hilbert might well have agreed with the starting point of relative realism but would have taken a more absolutist view of reality rather than subscribing to the `relative' side of the thesis. My view is that one does not have to swallow the philosophy for our approach to be useful. As with the philosophy of quantum mechanics, we do not need to get bogged down with what Real actually means as long as our model explains the perception of reality and key features of this perception at an operational level, which for us means principally its hierarchical structure whereby to some degree what we experience is determined by what assumptions we are working within. 

It also seems likely that Hilbert would not have agreed with the use to which we put this philosophical position. Where Hilbert might have wanted to build on empirical facts to find a complete set of precise axioms to shore up a theory more or less found by physicists, we are doing the opposite. In \cite{Ma:pri} we used our philosophical position to propose what we believe to be a single key principle or `crude axiom' (rather than a complete set of precise axioms) to help determine the mathematical structure of the not-yet-known theory of quantum gravity. Back in 1988 when I was finishing my thesis this would have been a necessity -- there was seemingly no real prospect of empirical tests for any theory of quantum gravity due to its energy scale being a factor of $10^{16}$ out of reach (a situation that is now  seen very differently). Secondly, theoretical physics back then was in my opinion still stumbling from structure to structure like Richard Feynman's drunk looking for their keys in the light of a streetlamp not because this is where he really thought he lost his keys but because this was the easiest place to look. Given our view that physics is a subset of mathematics, we apply model-building about the nature of physics to search for its mathematical structure a little more systematically in the space of mathematical structures. This is necessary  because in my opinion Nature is unlikely to use in the correct formulation of any ultimate theory only the mathematics already in math books (this being finite and limited by history and our collective imagination). Hence the search for the axioms of physics cannot be divorced from the discovery and structure of mathematics itself and needs to be reasonably consistent with broad features of the latter.   We now turn to what we proposed in \cite{Ma:pri} for this `axiom'. Since it's an axiom about the desired mathematical structure of physics, it is in some sense an axiom about the axioms of physics rather than an actual physical theory. Moreover, since we think of reality as having a kind hierarchical or fractal-like structure, it can apply at several different levels.

\section{Duality principle aka quantum Born reciprocity}\label{sec2}

If one steps back and looks at some general features of mathematics, one that stands out is the idea of a map or a function $f(x)$ being evaluated on an element $x$ in some space $X$ (the domain of $f$). This does  not cover all of mathematics but is common enough. If the space where the function takes values is something concrete and sufficiently self-evident that we think of it as directly observable (such as an integer or fraction or the real numbers obtained by completing fractions) then we can say that $f$ has a value at $x$ as some kind of measurement. Now the thing is that $f$ is also an element of a space of maps from $X$, let's call this $\hat X$, and who is to say that the number $f(x)$ is not actually $x(f)$, the value of $x$ at this point $f$ in $\hat X$? Mathematics itself has this striking duality between observer and observed running through it. Usually $X$ has some structure and we want $f$ to respect it, which in turn gives $\hat X$ some structure. In the dual point of view, we think of $X$ as $\hat{\hat{X}}$ and in the nicest cases this could be an identification of the two. The idea then is that if physics is a branch of mathematics, its central axiom should relate to the nature of what actually physics is. Physics is nothing if it is not the assumption that some structure is `out there' and that one can do an experiment to verify it. An experiment of course involves measuring or observing something, but actually what is an experimental fact is not something that really exists in isolation. Any experiment of any complexity usually involves a theory or some abstract relationship predicted by theory and from a theorists point of view an experiment `maps out' or confirms the assumed theory. In this sense an experiment represents a theoretical structure. From a dual point of view, however, an experimentalist might consider that their observations are self-evident facts and a theory merely compactly represents this data. For example, data points fall approximately on a line and the theorist represents this as a linear relationship. 
This suggests a kind of dualism in which either point of view should be equally correct as to which is the `real thing' and which is its set of representations. Clearly this is an idealisation or crude model of a much richer relationship but we see it as a key feature, as well as providing a  particular answer to the question posed by Plato's cave\cite{Ma:qg1}. 

\begin{figure} 
\[ \includegraphics[scale=.28]{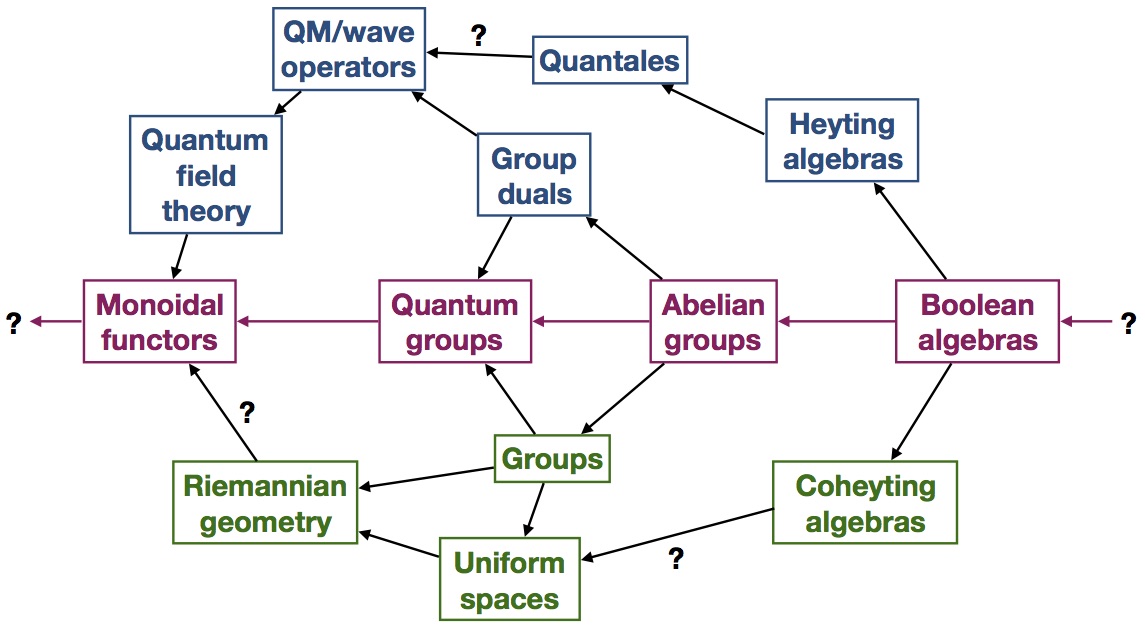}  \]
\caption{An axiom of Physics is the search for a self-dual structure in a self-dual category.\label{fig1}}
\end{figure}

Although we do not attempt to model the precise nature of theory and experiment across the physical sciences, one can  argue that something so basic {\em should} be reflected in the structure of any ultimate theory and meanwhile, we can see elements of it in particular contexts. We have written about this extensively elsewhere, so here we give only a concise overview.  First of all, we organise mathematics in general according to abstract concepts and their representations, as shown in  Figure~\ref{fig1} taken from \cite{Ma:pri,Ma:cat}. The arrows here are meant to be inclusion functors between categories of structures, loosely interpreted. The familiar case here is that of an Abelian group $G$. Its set of representations itself forms a group $\hat G$ and $G\subseteq \hat{\hat G}$ says that from mathematical point of view one is free to reverse which is the abstract structure and which is its representation. For example in Physics, $G$ could be position space $\R^n$ then $\hat G$ in a suitable setting would be momentum space $\R^n$, a self-dual example in the self-dual category. The {\em principle of representation-theoretic self duality}\cite{Ma:pri} or `generalised Mach principle' is the idea that Physics should admit a  reversal of which parts are structure and which parts are representation, for example which is position and which is momentum. This need not result in the same theory but merely a dual theory. The strong version is that the dual theory should have the same form but possibly with different values of parameters. From this point of view, Boolean algebra with its de Morgan duality is arguably the `birth' of physics\cite{Ma:cat}, while the next self-dual category beyond Abelian groups is Hopf algebras or `quantum groups'. Thus I argued in my 1988 PhD thesis that constructing noncommutative noncocommutative Hopf algebras could be seen as a toy model of constructing elements of quantum gravity, and used this  to obtain one of the two main classes of such true quantum groups at the time when these were first being introduced, the {\em bicrossproduct} ones associated to local Lie group factorisations. This was around the same time as V.G. Drinfeld introduced the other (and more famous) class of $q$-deformed quantum   groups coming from quantum integrable systems. I will say more about bicrossproducts shortly. 

\begin{figure}
\[\includegraphics[scale=.25]{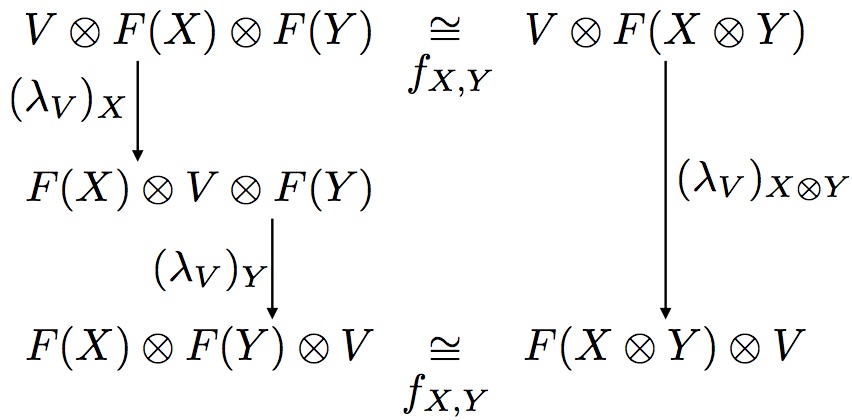}\]
\caption{Representation of a monoidal category underlying Drinfeld-Majid centre construction. \label{fig2}}
\end{figure}

Quantum groups here are a big enough category to include nonAbelian groups and their Fourier duality. If $G$ is a compact Lie group, say, its function algebra $C(G)$ and its group convolution algebra $C^*(G)$ can be completed to mutually dual Kac or Hopf-von Neumann algebras. At the algebraic level we have coordinate algebras $\C[G]$ and enveloping algebras $U(\cg)$ as essentially dual. Traditionally one has to do non-Abelian Fourier transform categorically but in the language of Hopf algebras it becomes quantum Fourier transform, for example $\C[G]\to \overline{U(\cg)}$ (indicating a suitable completion that includes exponentials). Here $U(\cg)$ is regarded as a `coordinate algebra' of a noncommutative space.  We will come to the physics of this shortly but for the moment we continue along the self-dual axis in Figure~\ref{fig1}. Here in the search for the `next' self-dual category, I found in 1990 the following duality construction $(\CC\to \CV)^\circ=\CC^\circ\to \CV$ for functors between monoidal categories\cite{Ma:rep,Ma:cat}. Here a monoidal category $\CC$ means there is a $\tens$ product which is associative up to an associator cocycle and $\CV$ another one, for example Vector Spaces, in which we construct our representations.  The objects of $\CC^\circ$ are pairs $(V,\lambda_V)$ where $V$ is an object of $\CV$ and $\lambda_V\in{\rm Nat}(V\tens F, F\tens V)$ is a natural transformation such that the diagram in Figure~\ref{fig2} commutes. Here $\lambda_V$ is a collection of morphisms $(\lambda_V)_X:V\tens F(X)\to F(X)\tens V$ for all $X\in \CC$ which are functorial in the sense of compatible with any morphisms $X\to Y$ in $\CC$ and the condition in the figure says that it `represents' the tensor product of $\CC$ as composition in $\CV$. Note that the monoidal functor $F$ comes equipped with an associated natural isomorphism $f$ in the sense of functorial isomorphisms $f_{X,Y}:F(X)\tens F(Y)\to F(X\tens Y)$ for all objects $X,Y$ in $\CC$, which we use. One has $\CC\subseteq \CC^\circ{}^\circ$ and the construction generalises both group and Hopf algebra duality. The tensor product of two `representations' is just 
\[ (\lambda_{V\tens W})_X=((\lambda_V)_X\tens\id)(\id\tens(\lambda_W)_X)\]
where we move $W$ past $F(X)$ then $V$ past $F(X)$. By a theorem of Mac Lane for monoidal categories, we suppress the associator between tensor products as these can be inserted afterwards. 

\begin{example} If $G$ is a finite group and $\CC$ the category of $G$-graded vector spaces, we can tensor product such spaces by the product of gradings in the group, obtaining a monoidal category. We take $F$ the functor that forgets the grading, then $\CC^\circ$ has as objects vector spaces $V$ equipped with natural isomorphisms $(\lambda_V)_X:V\tens F(X)\to F(X)\tens V$ sending $(\lambda_V)_X(v\tens x_g)=x_g\tens v\ra g$ for some right action of $G$ on $V$.  One can check that this meets the requirements above. Thus,  $\CC^\circ$ is essentially the category of representations of $G$. \end{example}

This connects our monoidal duality to non-Abelian group duality. The latter also includes Hopf algebra duality when appropriately formulated. A genuinely new example of considerable interest these days in topological quantum field theory is the following.

\begin{example} (Drinfeld-Majid centre.) A special case of the above construction is when $\CV=\CC$ and $F={\rm id}$. This case was found independently by Drinfeld according to a private communication cited in \cite{Ma:rep} but the definitions and proofs are identical to the $\CC^\circ$ construction, just leaving out the symbol $F$. In this case there is a tautological braiding if we assume the $\lambda$ are isomorphisms. \end{example}

Drinfeld's private communication here was a letter to me after he came across the preprint version of \cite{Ma:rep}. Hence it is not the case that $\CC^\circ$ was obtained as a generalisation of Drinfeld's work as sometimes assumed, we simply came to essentially the same construction for different reasons. My reason was the principle of representation-theoretic self-duality while Drinfeld's was to generalise his famous double construction for quantum groups in \cite{Dri}. Some of my follow-up work was \cite{Ma:cat}.

How is this dualism reflected in Physics? One setting already alluded to and which we have called {\em quantum Born reciprocity (QBR)}, is Fourier duality between position and momentum space and its generalisations. In the Abelian group case this is just wave particle duality, but it also works in the nonAbelian case. If the universe is spatially $S^3=SU_2$ (and it might be) then spatial momentum is the representations of this. These form a category but as noted one can also see Fourier transform at the Hopf algebra level from $\C[SU_2]$ to (a completion of) $U(su_2)$ where the latter is regarded as a quantum momentum space $[p_i,p_j]={\imath \hbar \over l_c }\epsilon_{ijk}p_k$. Here $l_c$ is the cosmological curvature scale and $p_i$ are left-invariant vector fields. Dually, if the momentum space of some system were to be the nonAbelian group $SU_2$ then the Fourier dual would be the quantum spacetime with relations $[x_i,x_j]=\imath\lambda_P\epsilon_{ijk}x_k$ where $\lambda_P$ is a length scale and the $U(su_2)$ generators are now regarded as position coordinates. This as we will see shortly is thought to be the case in some models of 3D quantum gravity. 

\begin{figure}
\[\includegraphics[scale=.35]{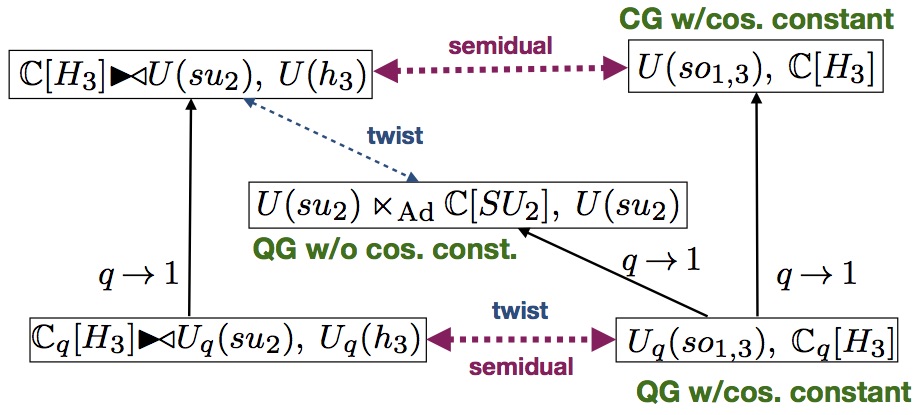}\]
\caption{Quantum Born reciprocity expressed as semidualisation in 3D quantum gravity and its limits.\label{fig3}}
\end{figure}

In my PhD thesis\cite{Ma:pla,Ma:phy} I took this point of view and the above self-duality principle as a motivation to look for self-dual type Hopf algebras, and constructed these in the `bicrossproduct' form $\C[M]\bicross U(\cg)$ with dual $U(\cm) \cobicross \C[G]$ associated to any local factorisation of a Lie group $X=G\bowtie M$. These were originally thought of these as quantum phase space but since 1994 in \cite{MaRue} I have also thought of them as quantum Poincar\'e groups acting on $U(\cm)$ and $U(\cg)$ respectively as auxiliary quantum spacetimes. In fact there are different covariant systems with equivalent data related by {\em semidualisation}\cite{Ma:pla,Ma:phy} (where one Hopf algebra is systematically replaced by its dual),
\[ \C[M]\bicross U(\cg),\ U(\cm)\Leftrightarrow  U(\cx),\ \C[M];\quad U(\cm) \cobicross \C[G],\ U(\cg)  \Leftrightarrow U(\cx),\ \C[G]\]
where $\cx$ is the Lie algebra of $X$ and in each pair we give the (possibly quantum) symmetry group and the (possibly quantum) spacetime algebra on which it acts. The relevant factorisation for 3D quantum gravity is $SL(2,\C)=SU_2\bowtie H_3$ where $H_3=\R^2\rtimes\R$ is the group of upper-triangular matrices in the Iwasawa decomposition. Focusing on the first two pairs, we have the top line of Figure~\ref{fig3} where on the top left $U(h_3)$ is the quantum spacetime
\begin{equation}\label{MR} [x_i,t]=\imath\lambda_P x_i\end{equation}
for $i=1,2$ which is the 3D version of the Majid-Ruegg quantum spacetime \cite{MaRue}. In its Poincar\'e quantum group the momentum is commutative because its `enveloping algebra' is the commutative Hopf algebra $\C[H_3]$ but curved as $H_3$ is non-Abelian. The semidual of this on the top right is a classical model of a particle on  $H_3$ as curved position space with its classical $U(so_{1,3})=U(su_2)\bowtie U(h_3)$ symmetry containing $U(h_3)$ as the translational  momentum. So the roles of position and momentum are swapped between the two models -- an example of QBR. It is also striking that the  model on the right is classical (a particle on a curved space $H_3$) while the other model is quantum, so QBR interchanges classical and quantum. 

In fact this picture $q$-deforms \cite{MaSch} as we show on the bottom line of Figure~\ref{fig3}, where the model on the bottom right is thought to encode quantum gravity with cosmological constant via an expression of the form $q=e^{-\lambda_P/l_c}$. Its QBR-dual model on the bottom left when $q\ne 1$ is at some algebraic level isomorphic to two copies of $\C_q[H_3]=U_q(su_2)^{cop}$ acting on $U_q(h_3)=\C_q[SU_2]^{op}$ up to some technicalities, i.e. a classical but $q$-deformed particle on a 3-sphere, and this is related by a categorical equivalence (a Drinfeld twist) to  the model on the right. In other words, 3D quantum gravity with cosmological constant is in some sense {\em self-dual} under QBR. Finally, we can take $q\to 1$ in different ways and the first one, on the outer right, is $\lambda_P\to 0$ (so a classical but curved position space). Alternatively we can send $l_c\to \infty$ and this is the model in the centre of the figure encoding 3D quantum gravity without cosmological constant (to see this one should write $U_q(so_{1,3})=U_q(su_2)\bowtie U_q(h_3)=U_q(su_2)\bowtie\C_q[SU_2]^{\rm op}$ up to some technicalities, and then take the limit). The diagonal twist equivalence between this conventional version of 3D quantum gravity with $U(su_2)$ quantum spacetime and the one we began with (on top left) was recently shown by P. Osei and the author\cite{MaOse}. More details of the point of view for 3D quantum gravity are in \cite{MaSch}.

If we leave the self-dual axis then the dual structure is not of the same type but is still a structure. If we take a more categorical view of group duals then the dual of a nonAbelian Lie group generally comes down to quantisation, such as the method of coadjoint orbits or the construction of the relevant representations of the Poincar\'e group in flat spacetime by particle-waves of different mass and spin. The dual object consists of all of these considered together. Geometrically, we are diagonalising natural Laplacians or wave operators. The same principle applies when the spacetime is a compact Lie group, the particle content is closely related to its representation theory. Compact Lie groups on the other hand are the simplest examples of Riemannian manifolds and we can similarly albeit more loosely think of  the `dual'  of the latter as coming down to quantum mechanics or wave operators of different types. Recall that quantum mechanics in nice cases can be seen as the non-relativistic limit of the Klein-Gordon spacetime Laplacian for fields where $e^{-\imath m c^2 t/\hbar}$ is factored out. In fact a Riemannian or pseudo-Riemannian manifold cannot be totally reconstructed from the scalar Laplacian alone but if we use the Dirac operator then one has  Connes' reconstruction of a spin manifold from an abstractly defined `spectral triple' in the commutative case\cite{Con:cmp}. Either way, if one extends these ideas from classical to quantum field theory then logically an aspect of this should be that it corresponds to {\em quantum Riemannian geometry}, where spacetime coordinates become noncommutative. This should then be taking us towards the self-dual axis as shown Figure~\ref{fig1}. Interestingly, \cite{Fred} have now constructed quantum field theories on  curved spacetimes as a functors from the monoidal category of globally hyperbolic spacetimes to noncommutative $C^*$-algebras with tensor product, which is some kind of arrow from QFT to monoidal functors in Figure~\ref{fig1}. Such functors are not self-dual under $(\ )^\circ$ but then this is not yet quantum gravity. At any rate, one could speculate in this context that Einstein's equation might eventually emerge as the classical limit of a self-duality condition as it equates the Einstein tensor from the geometry side to the expectation value of the stress energy tensor from the quantum field theory side, probably requiring both to be generalised so as to be in a self-dual setting.  I do not know the final framework for this but the monoidal category dual may be a step in the right direction. In physical terms, the self-dual nature of quantum gravity as we see it is in line with the fact that the metric is the background geometry while its fluctuations are spin-2 fields and hence among the `representations' of the background geometry in the loose sense discussed above. 

In summary, we were led on philosophical grounds to the view that spacetime should be both curved and `quantum' in the sense of a noncommutative coordinate algebra, as a consequence of a deep self-duality principle for quantum gravity\cite{Ma:pri,Ma:spbook,Ma:qg1}. We call this aspect the {\em quantum spacetime hypothesis}. It's a prediction which, if confirmed, would be on a par with the discovery of gravity and indeed dual to it. What it could entail mathematically is our next topic.

\section{Axioms of quantum Riemannian geometry}\label{sec3}

I will recap a constructive approach to this from my own work in the last decade (much of it with Edwin Beggs) rather than the better-known `Dirac operator' approach of Connes\cite{Con} expressed in the axioms of a spectral triples. The two approaches have recently begun to come together with our programme of {\em geometric realisation} of Connes' spectral triples\cite{BegMa:spe}. 

In fact all main approaches have in common (in our case as starting point) the notion of differential forms $(\Omega,\extd)$ over a possibly noncommutative algebra $A$  as a differential graded algebra. This means 
\[\Omega=\oplus_n\Omega^n,\quad \extd:\Omega^n\to \Omega^{n+1},\quad \extd^2=0\]
where $\Omega^0=A$ and $\extd$ obeys a graded-Leibniz rule with respect to the graded product $\wedge$. We assume that $\Omega$ is generated by $A,\Omega^1$ in which case one may focus on $(\Omega^1,\extd)$ first and construct higher differential forms as a quotient of the tensor algebra of this over $A$. Here 
\[\extd:A\to \Omega^1,\quad  \extd(ab)=(\extd a)b+a(\extd b),\quad  (a\extd b)c=a((\extd b)c)\]
where $\Omega^1$ has an associative multiplication from the left and the right by $A$ (one says that $\Omega^1$ is an $A$-bimodule) and $\extd$ is a derivation.

The next ingredient is a left connection, 
\[\nabla :\Omega^1\to \Omega^1\tens_A\Omega^1,\quad \nabla(a\omega)=\extd a\tens \omega+ a\nabla \omega\]
which is a {\em bimodule connection}\cite{DubMic,Mou} if there also exists a bimodule map $\sigma$ such that
\[ \sigma:\Omega^1\tens_A\Omega^1\to \Omega^1\tens_A\Omega^1,\quad \nabla(\omega a)=(\nabla\omega)a+\sigma(\omega\tens\extd a).\]
The map $\sigma$ if it exists is unique, so this is not additional data but a property that some left connections have. 
In the classical case where $A=C^\infty(M)$, if $X$ is a vector field then a connection $\nabla$ defines a covariant derivative $\nabla_X:\Omega^1\to \Omega^1$ by evaluating $X$ against the left output of $\nabla$ (this also works with care in the quantum case). However, we consider all such covariant derivatives together by leaving $\nabla$ as a 1-form valued operator on 1-forms.  The curvature and torsion of a left connection, see for example \cite{Ma:rie}, are 
\[  R_\nabla=(\extd\tens\id-\id\wedge\nabla)\nabla:\Omega^1\to \Omega^2\tens_A\Omega^1,\quad T_\nabla=\wedge\nabla-\extd:\Omega^1\to\Omega^2.\]

Incidentally, all the same ideas except for the torsion hold for any vector bundle, which we axiomatize via its space of sections $E$ as a left module over $A$,  and $\nabla_E:E\to \Omega^1\tens_AE$. In the bimodule case, if $E,F$ are bimodules and $\nabla_E,\nabla_F$ are bimodule connections then the tensor product $E\tens_A F$ has a bimodule connection
\[\nabla_{E\tens F}(e\tens f)=\nabla_E e\tens f+ (\sigma_E\tens\id)(e\tens\nabla_F f)\]
and a certain $\sigma_{E\tens F}$. This makes the collection of such pairs $(E,\nabla_E)$ into a monoidal category  with morphisms usually taken as bimodule maps that intertwine the connections.  There is a forgetful functor from this to the category of bimodules over $A$, so this is an example of  monoidal functor (in the sense of Figure~\ref{fig1}) associated to any manifold  and to any algebra more generally. 

Next we consider a Riemannian metric $g=g^1\tens_A g^2\in \Omega^1\tens_A\Omega^1$ (sum of such terms understood). We want to this to non-degenerate in the sense that there exists a bimodule map $(\ ,\ ):\Omega^1\tens_A\Omega^1\to A$ that is inverse, $(\omega,g^1)g^2= \omega=g^1(g^2,\omega)$ 
for all $\omega\in\Omega^1$. In this case $(\omega, ag^1)g^2=(\omega a, g^1)g^2=\omega a=(\omega,g^1)g^2a$ tells us that $[a,g]=0$ for all $a$, i.e. $g$ has to be central\cite{BegMa:cqg}. So even though we are doing noncommutative geometry and do not assume that 1-forms commute with functions, we will need the metric to be central. This is a significant constraint on quantum spacetime in the noncommutative case which is invisible classically. We also usually want the metric to be quantum symmetric in the sense $\wedge(g)=0$.

Finally, we want $\nabla$ to be metric compatible. There is a weak notion that makes sense for any left connection, namely it is `weak quantum Levi-Civita' if it is torsion free and
\[ coT_\nabla=(\extd\tens\id-\id\wedge\nabla)g\in \Omega^2\tens_A\Omega^1\]
vanishes. This `cotorsion tensor' was introduced in \cite{Ma:rie} and classically the condition says that $\nabla_\mu g_{\nu\rho}-\nabla_\nu g_{\mu\rho}=0$. In the case of a bimodule connection we can do better and we say this is {\em quantum Levi-Civita (QLC)} if it is torsion free and $\nabla g=0$ where $\nabla$ extends to $\Omega^1\tens_A\Omega^1$ by the tensor product formula. 

Usually one wants $A$ to be a $*$-algebra and for $*$ to be extended as a graded-involution to $\Omega$ commuting with $\extd$, and for $g^\dagger=g$, $\nabla\circ *= \sigma\circ\dagger\circ \nabla$ where $(\omega \tens_A\eta)^\dagger=\eta^*\tens_A \omega^*$. These reality conditions in a self-adjoint basis (if one exists) would ensure that the metric and connection coefficients  are real at least in the classical limit. This completes our lighting review. 

By now there are many specific quantum Riemannian geometries constructed, for example on the quantum sphere $\C_q[S^2]$, see \cite{Ma:spin}, on the quantum spacetime (\ref{MR}), see \cite{BegMa:cqg},  on the functions on the permutation group $\C(S_3)$,  and on its dual $\C S_3$, see \cite{MaTao}, each with natural differential structure, quantum metric and QLC's or weak QLCs according to the model. The Ricci tensor is only partially understood because to follow the usual trace contraction one would need a lifting map $i:\Omega^2\to \Omega^1\tens_A\Omega^1$, which is an additional datum. The Dirac operator is also only partially understood needing both a `spinor' bundle with connection compatible with $\nabla$ and a `Clifford action'. At least for $\C_q[S^2]$ one can come close to the axioms of a Connes spectral triple at least at an algebraic level before any functional analysis completions\cite{BegMa:spe}. 

\section{Poisson-Riemannian geometry}\label{sec4}

Our motivation has been that quantum geometry deforms classical geometry by order $\lambda_P$ corrections, as a measure of some quantum gravity effects. The semiclassical theory of which the above is a quantisation was recently  worked out by Beggs and the author in \cite{BegMa:prg}. This theory is to aspects of quantum gravity as classical mechanics is to quantum mechanics, except the deformation parameter is not $\hbar$, so a kind of `classical quantum gravity'. One could imagine other applications,  including to quantum mechanics if the phase space also has a natural Riemannian structure, so we will just call the parameter $\lambda$ (and take conventions where it is imaginary).

The first layer of this is of course the Poisson structure\index{Poisson structure}, a tenet of mathematical physics since the early days of quantum mechanics being to `quantise' functions  $C^\infty(M)$ on a manifold to a noncommutative algebra $A$. We suppose  that 
\[ a\bullet  b= ab + O(\lambda)\]
where we denote the $C^\infty(M)$ product by juxtaposition and the $A$ product by $\bullet$.
We assume all expressions can be expanded in $\lambda$ and equated order by order. In this case 
\[ a\bullet  b-b\bullet  a=\lambda\{a,b\} + O(\lambda^2)\]
defines a map $\{\ ,\ \}$ and the assumption of an associative algebra quickly leads to the necessary feature that this is a Lie bracket (i.e.\ antisymmetric and satisfies the Jacobi identity\index{Jacobi identity}, making $C^\infty(M)$ into a Lie algebra) and  $\hat a:=\{a, \}$ is a (`Hamiltonian') vector field associated to a function $a$. It is known that every such Poisson bracket can be quantised to an associative algebra at least at some formal level\cite{Konts}. The second layer is to find a differential structure $\Omega^1$ deforming the classical $\Omega^1(M)$. One can similarly analyse the data  for this by defining a map $\nabla$ by
\[ a\bullet  (\extd b)-( \extd b)\bullet  a=\lambda \nabla_{\hat a}\extd b + O(\lambda^2).\]
The assumption of a left action and the Leibniz rule for $\extd$, requires at order $\lambda$ that 
\begin{equation}\label{poicomp} \nabla_{\hat a}(b\,\extd c)=\{a,b\}\extd c+ b\,\nabla_{\hat a}\extd c,\quad 
\extd\{a,b\}=\nabla_{\hat a}\extd b- \nabla_{\hat b}\extd a\end{equation}
(these follow easily from $[a, b\bullet\extd c]=[a,b]\bullet\extd c+ b\bullet[a,\extd c]$ and $\extd [a,b]=[\extd a,b]+[a,\extd b]$). The first condition of (\ref{poicomp}) says that $\nabla$ is a covariant derivative along Hamiltonian vector fields $\hat a$ and the second is an additional `Poisson-compatibility'. The first part of (\ref{poicomp}) applies similarly for any bundle and can be formulated as $\nabla$ a contravariant or Lie-Rinehart connection\cite{Heu}, while the second part was observed in \cite{Haw,BegMa:sem}. Finally, the associativity of left and right actions on a bimodule gives
\[ R_\nabla(\hat a,\hat b):=\nabla_{\hat a}\nabla_{\hat b}-\nabla_{\hat b}\nabla_{\hat a}-\nabla_{\widehat{\{a,b\}}}=0\]
(this follows from the Jacobi identity $[a,[\extd b,c]]+[\extd b, [c,a]]+[c,[a,\extd b]]=0$).  So a zero curvature Poisson-compatible partially-defined connection is what we strictly need.  

In \cite{BegMa:prg} we make two convenient variations.  First of all we are {\em not} going to require zero curvature because the effect of curvature is visible only at order $\lambda^2$, so we do not really need this in the order $\lambda$ theory. If there is curvature then it will not be possible to have an associative differential calculus of classical dimension on $A $ but this is actually a situation that one frequently encounters in noncommutative geometry. We can either absorb this in a higher dimensional associative differential structure or we can live with nonassociative differentials at order $\lambda^2$. Strictly speaking,  the same applies to the Poisson bracket obeying the Jacobi identity not being strictly needed at order $\lambda^2$ in which case we would have $A $ itself being non-associative.  Secondly, for simplicity, we are going to make the assumption that $\nabla_{\hat a}$ is indeed the restriction of an actual connection $\nabla$. This will allow us to speak more freely of geometric concepts such as the contorsion tensor. In fact this assumption is not critical; if the Poisson tensor in these coordinates is $\omega^{\mu\nu}$ then we are in most formulae making use only of the combination $\nabla^\mu:=\omega^{\mu\nu}\nabla_\mu$ rather than the full covariant derivatives $\nabla_\mu$ themselves. It means that our data has redundant `auxiliary modes' that do not affect the quantum differential structure at order $\lambda$, a situation not unfamiliar from other situations such as gauge theory. There is also the matter of extending from $\Omega^1$  to forms of all degree but this turns out\cite{BegMa:prg} to impose no further conditions. 

The third layer is the construction of a quantum metric and the natural data for this will be a classical metric $g$ on $M$. As one might guess the metric compatibility of $\nabla$ is just that $\nabla g=0$. 
To avoid confusion we will write $\widehat\nabla$ for the classical Levi-Civita connection of $g$ and we let $S$ be the contorsion tensor of $\nabla$ whereby  $\widehat\nabla=\nabla+S$. It is well-known in General Relativity that a metric compatible connection is determined by its torsion tensor $T$ or equivalently a cotorsion tensor $S$  antisymmetric in its outer indices when all indices are lowered. Hence under our simplifying assumption the data for $\nabla$ can be thought of as $T$ or $S$. In this case Poisson compatibility of $\nabla$ can be written as\cite{BegMa:prg},
\begin{equation}\label{nablaS} \widehat{\nabla}_\gamma \omega^{\alpha \beta}+S^\alpha{}_{\delta \gamma}\omega^{\delta \beta}+S^\beta{}_{\delta \gamma}\omega^{\alpha \delta}=0. \end{equation}
The fourth layer is more specialised as it is specifically the quantisation data for  a bimodule quantum Levi-Civita connection (one could be happy with something weaker) and comes down to the identity
\begin{equation}\label{levicond} \widehat\nabla_\rho\CR_{\mu\nu}+S^\beta{}_{\alpha \nu}H^{\alpha}{}_{\beta\rho\mu}-S^\beta{}_{\alpha\mu}H^{\alpha}{}_{\beta\rho\nu}=0\end{equation}
where the curvature $R$ of $\nabla$ combines with the contorsion to define
\begin{equation}\label{HR} H^{\alpha}{}_{\beta\mu\nu}=g_{\beta\gamma}\omega^{\gamma\rho}\left(\nabla_\rho S^\alpha{}_{\mu\nu}+R^\alpha{}_{\nu\mu\rho}\right),\quad \CR_{\mu\nu}={1\over 2}\left(H^{\alpha}{}_{\alpha\mu\nu}-H^{\alpha}{}_{\alpha\nu\mu}\right).\end{equation}
The latter is called the {\em generalised Ricci 2-form} associated to our classical data. In summary, the field equations of Poisson-Riemannian geometry come down to\cite{BegMa:prg}: 

\begin{enumerate}
\item[(0)]  A metric $g_{\mu\nu}$ and an antisymmetric bivector $\omega^{\mu\nu}$ typically obeying the Poisson bracket Jacobi identity; 

\item[(1)] A metric compatible connection $\nabla$ at least along Hamiltonian vector fields;

\item[(2)] Poisson-compatibility of $\nabla$ given in the fully defined case by (\ref{nablaS});

\item[(3)] The optional quantum Levi-Civita condition (\ref{levicond}).
\end{enumerate}

These equations can be quite restrictive, particularly if one also wants to preserve a symmetry. 

\begin{example} (Quantizing the Schwarzschild black hole\cite{BegMa:prg}) Solving the above equations for the Schwarzschild metric in  polar coordinates $t,r,\theta,\phi$, and asking to preserve rotational symmetry leads to a unique Poisson tensor $\omega$ and unique $\nabla$ up to auxiliary modes. This leads to $r,t,\extd r,\extd t$ central (unquantised) and for each $r,t$ one has a radius $r$ `nonassociative fuzzy sphere' 
\[ [z_i,z_j]=\lambda\eps_{ij}{}_k z_k,\quad  [z_i,\extd z_j]=\lambda z_j \eps_i{}_{mn} z_m\extd z_n.\]
to order $\lambda$ in coordinates where $\sum_i z_i^2=1$. Here $\nabla$ on $S^2$ is the Levi-Civita connection with constant curvature, hence $\Omega^1$ is not associative at order $\lambda^2$.
\end{example}

This uniqueness result was extended to generic static spherically symmetric spacetimes in \cite{FriMa}. 

\section{Quantum gravity on a square graph}  \label{sec5}

The mathematics of quantum Riemannian geometry is simply more general than classical Riemannian geometry and includes discrete\cite{Ma:gra} as well as deformation examples. What is significant in this section is that whatever we find {\em emerges} from little else but the axioms applied to a square graph as `manifold'. 

Let $X$ be a discrete set and $A=\C(X)$ the usual commutative algebra of complex functions on it as our `spacetime algebra'. It is an old result that its possible 1-forms and differential $(\Omega^1,\extd)$ are in 1-1 correspondence with directed graphs with $X$ as the set of vertices. Here $\Omega^1$ has basis $\{\omega_{x\to y}\}$ over $\C$ labelled by the arrows of the graph and differential $\extd f=\sum_{x\to y}(f(y)-f(x))\omega_{x\to y}$. In this context a quantum metric
\[ g=\sum_{x\to y}g_{x\to y}\omega_{x\to y}\tens\omega_{y\to x}\in \Omega^1\tens_{C(X)}\Omega^1\]
requires weights $g_{x\to y}\in \R\setminus\{0\}$ for every edge and for every edge to be bi-directed (so there is an arrow in both directions). Taking all weights to be 1 is the so-called `Euclidean metric'  \cite{Ma:gra}.  The calculus over $\C$ is compatible with complex conjugation on functions $f^*(x)=\overline{f(x)}$ and $\omega_{x\to y}^*=-\omega_{y\to x}$. Finding a QLC for a metric depends on how $\Omega^2$ is defined and one case where there is a canonical choice of this is $X$ a group and the graph a Cayley graph generated by right translation by a set of generators. Previously a QLC was found for some specific groups and the `Euclidean metric' but here we give a first calculation for a reasonably general class of metrics.  

\begin{figure}\[ 
\includegraphics[scale=.18]{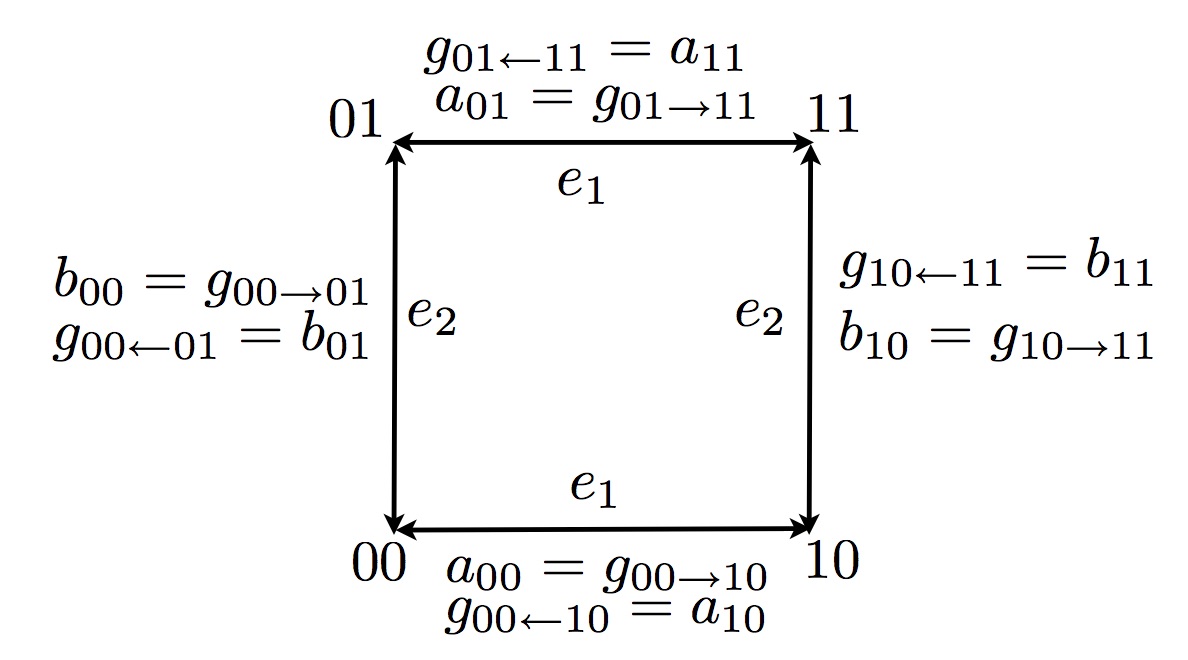}\]
\caption{Metric coefficients $a,b$ are interpreted as arrow lengths in our formulation of discrete gravity on a square. \label{fig4}}\end{figure}

We take $X=\Z_2\times\Z_2$ with its canonical 2D calculus given by a square graph with vertices $00,01,10,11$ in an abbreviated notation as shown in Figure~\ref{fig4}. The graph is regular and there are correspondingly two basic 1-forms 
\[ e_1=\omega_{00\to 10}+\omega_{01\to 11}+\omega_{10\to 00}+\omega_{11\to 01},\quad e_2=\omega_{00\to 01}+\omega_{10\to 11}+\omega_{01\to 00}+\omega_{11\to 10}\] with relations and differential 
 \[ e_i f=(R_if)e_i,\quad \extd f=(\del^1 f)e_1+(\del^2 f)e_2\]
where $R_1f$ shifts by 1 mod 2 (i.e. takes the other point) in the first coordinate, similarly for $R_2$, and  $\del^i=R_i-\id$.   The exterior algebra is the usual Grassmann algebra on the $e_i$ (they anticommute). The general form of a quantum  metric and its inverse are 
\[  g=a e_1\tens e_1+b e_2\tens e_2,\quad (e_1,e_1)={1\over R_1 a},\quad (e_2,e_2)={1\over R_2 b},\quad (e_1,e_2)=(e_2,e_1)=0\]
where the $a,b$ are non-vanishing functions. With the standard $*$ structure $e_i^*=-e_i$, the metric obeys the reality condition in Section~\ref{sec3} if $a,b$ are real valued. In terms of the graph their 8 values are equivalent to the values of $g$ on the 8 arrows as shown in Figure~\ref{fig4}. It is natural here to focus on the symmetric case where the metric weight assigned to an edge does not depend on the direction of the arrow. This means $\del^1 a=\del^2b=0$ and we assume this now for simplicity. In  this case we find a 1-parameter family of torsion free metric compatible or `quantum Levi-Civita' connections:
\[ \nabla e_1=(1+Q^{-1})e_1\tens e_1+(1-\alpha)(e_1\tens e_2+e_2\tens e_1)- {b\over a}(R_2\beta-1)e_2\tens e_2,\]
\[ \nabla e_2=-{a\over b}(R_1\alpha-1)e_1\tens e_1+(1-\beta)(e_1\tens e_2+e_2\tens e_1)+(1-Q)e_2\tens e_2,\]
\[ \sigma(e_1\tens e_1)=-Q^{-1}e_1\tens e_1+{b (R_2\beta-1)\over a}e_2\tens e_2,\quad \sigma(e_2\tens e_2)=Qe_2\tens e_2+{a (R_1\alpha-1)\over b}e_1\tens e_1,\]
\[ \sigma(e_1\tens e_2)=\alpha e_2\tens e_1+(\alpha-1)e_1\tens e_2,\quad \sigma(e_2\tens e_1)=\beta e_1\tens e_2+ (\beta-1)e_2\tens e_1\]
where $Q,\alpha,\beta$ are functions on the group defined as
\[ Q=(q,q^{-1},q^{-1},q)=q^\chi,\quad \alpha=({a_{01}\over a_{00}}, 1, 1, {a_{00}\over a_{01}}),\quad \beta=(1, {b_{10}\over b_{00}},  {b_{00}\over b_{10}},1)\]
when we list the values on the points in the same binary sequence as above. Here $q$ is a free parameter and $\chi(i,j)=(-1)^{i+j}=(1,-1,-1,1)$ is a function on $\Z_2\times\Z_2$.  If we write $\sigma$ as a matrix $\sigma^{i_1i_2}{}_{j_1j_2}$ where the multi-indices are in order $11,12,21,22$, is
\begin{equation}\label{sigma8v} \sigma=\begin{pmatrix} -Q^{-1} & 0 & 0 & {a(R_1\alpha-1)\over b}\cr 0 & \alpha-1 & \beta & 0\cr 0 & \alpha &\beta-1 & 0\cr {b(R_2\beta-1)\over a} & 0 & 0& Q\end{pmatrix} . \end{equation}
What we have coming out of the axioms is a field of such `generalised braiding' matrices because the entries here are functions on the group. The eigenvalues are  $-1, \alpha\beta, -Q^{-1}, Q$ as functions on the group.  Notice that these `generalised braidings' have a broadly  8-vertex form normally associated with quantum integrable systems but here arising  out of nothing but the requirements of quantum Riemannian geometry applied to a square graph.

The Laplacian for the above QLC's are computed as
\begin{eqnarray*} \Delta f=(\ ,\ )\nabla(\del_i f e_i)=-{2\over a}\del_1 f-{2\over b}\del_2 f+\del_i f(\ ,\ )\nabla e_i=\left({Q^{-1}-R_2\beta\over a}\right)\del_1 f- \left({Q+R_1\alpha\over b}\right)\del_2 f\end{eqnarray*}
using our formula for $\nabla$, the connection property, and $\del_i^2=-2\del_i$. The curvatures are given by
\begin{align*} R_\nabla e_1=&\left(Q^{-1}R_1\alpha-Q\alpha+(1-\alpha)(R_1\beta-1)+{R_2 a\over a}(R_2\beta-1)(R_2R_1\alpha-1)\right){\rm Vol}\tens e_1\\
&+
\left(Q^{-1}(1-\alpha)+\alpha(R_2\alpha-1)+Q^{-1}{R_1b\over a}(\beta^{-1}-1))+{b\over a}(R_2\beta-1)R_2\beta\right){\rm Vol}\tens e_2\end{align*}
where ${\rm Vol}=e_1\wedge e_2$, and a similar formula for $R_\nabla e_2$ interchanging $e_1, e_2$;  $R_1,R_2$;  $\alpha,\beta$; $a,b$ and $Q,-Q^{-1}$ (so that ${\rm Vol}$ also changes sign). One can discern contributions from $q\ne 1$ and from $a,b$ non-constant. The connection reality condition comes down to
\begin{equation}\label{qZ2Z2}|q|=1\end{equation}
so that in particular the function $Q-Q^{-1}$ is pointwise imaginary. 

Next we find the Ricci tensor defined by a lifting map $i$, for which in our case there is a canonical choice   $i({\rm Vol})={1\over 2}(e_1\tens e_2-e_2\tens e_1)$. If we write $R_\nabla e_i=\rho_{ij}{\rm Vol}\tens e_j$ then 
\[ {\rm Ricci}=((\ ,\ )\tens\id)(\id\tens i\tens\id)(\id\tens R_\nabla)(g)={1\over 2}\begin{pmatrix}-R_2\rho_{21} & -R_2\rho_{22}\\ R_1\rho_{11} & R_1\rho_{12}\end{pmatrix}\]
as the matrix of coefficients on the left in our tensor product basis. Applying $(\ ,\ )$ again, we have scalar curvature
\[ S={1\over 2 }\left( -  {R_2\rho_{21}\over a}+{R_1 \rho_{12}\over b}\right)\]
which is invariant under the interchange above. For the `purely quantum' case where $q\ne 1$ and $a,b$ are constant, the QLCs and their curvature reduce to 
\[ \nabla e_1=(1+Q^{-1})e_1\tens e_1,\quad \nabla e_2=(1-Q)e_2\tens e_2,\]
\[ R_\nabla e_1=-(Q-Q^{-1}){\rm Vol}\tens e_1,\quad R_\nabla e_2=(Q-Q^{-1}){\rm Vol}\tens e_2\]
as the intrinsic quantum Riemannian geometry of $\Z_2\times\Z_2$ with its square metric. This has
\[ {\rm Ricci}={Q-Q^{-1}\over 2}(e_1\tens e_2+e_2\tens e_1),\quad S=0\]
which we see is quantum symmetric but does not obey the same reality condition as the metric if we impose (\ref{qZ2Z2}) needed for the connection to obey its `reality' condition. This is a purely quantum effect since classically there would be no curvature when $a,b$ are constant. 

The general Ricci curvature is quite complicated but for $q=1$, say, it has  values
\begin{equation}\label{canRZ2Z2} {\rm Ricci}_{q=1}={1\over 2}\begin{pmatrix}{1\over b}(-{\del_2 a\over\alpha}+\chi {\del_1 b\over\beta}) & -{\del_1 b\over b}(\alpha+{1\over\alpha}-\chi-2)   \\ 
-{\del_2 a\over a}(\beta+{1\over\beta}-\chi-2)   & {1\over a}(-{\del_2 a\over\alpha}+\chi {\del_1 b\over\beta})\end{pmatrix}\end{equation}
for the matrix of coefficients. This is not quantum symmetric or real in the sense of the metric. For the scaler curvature the general formula is
\[ S=-{1\over 4 ab}\left((3+q+(1-q)\chi){\del_2 a\over\alpha}+(1-q^{-1}-(3+q^{-1})\chi) {\del_1 b\over\beta}\right).\]
Finally, it is not obvious what measure we should use to integrate either of these but if we take measure $\mu=|ab|=ab$ (we assume for now the $a,b$ are positive edge lengths, i.e. the theory has Euclidean signature) and sum over $\Z_2\times\Z_2$ then we have 
\begin{equation}\label{EH} \int S=\sum_{\Z_2\times\Z_2} \mu S=(a_{00}-a_{01})^2({1\over a_{00}}+{1\over a_{01}})+(b_{00}-b_{10})^2({1\over b_{00}}+{1\over b_{10}}).\end{equation}
independently of $q$.  We consider this action as some kind of energy of the metric configuration. If we took other measures such as $\mu=1$ or $\mu=\sqrt{|g|}=\sqrt{|ab|}$ then we would not have invariance under $q$ so the action would not depend only on the metric but on the choice of $\nabla$.

Next we Fourier transform on $\Z_2\times\Z_2$ to write our results in `momentum space'. We have 
\[ 1,\quad \phi(i,j)=(-1)^i=(1,1,-1,-1),\quad \psi(i,j)=(-1)^j=(1,-1,1,-1),\quad \phi\psi=\chi\]
\[\del_1\phi=-2\phi,\quad \del_2\phi=0,\quad \del_1\psi=0,\quad \del_2\psi=-2\psi \]
as the plane waves and given the conditions we imposed on $a,b$, we can expand these in terms of four real momentum space coefficients as
\[a=k_0+k_1\psi,\quad b=l_0+l_1\phi.\]
Then some computation gives the Scalar curvature for $q=1$ as
\begin{align*}S=\frac{2}{(k_0^2-k_1^2)  (l_0^2-l_1^2)}&\Big((l_0-l_1) (k_1 (k_0+k_1)-l_1 (k_0-k_1)),(k_0+k_1) (l_1 (l_0+l_1)-k_1 (l_0-l_1)),\\ &(k_0-k_1)(k_1 (l_0+l_1)-l_1 (l_0-l_1)),(l_0+l_1) (l_1 (k_0+k_1)-k_1 (k_0-k_1))\Big).\end{align*}
With measure $\mu=ab$ as above, this gives
\[ \int S=8\left(  {k_0k_1^2\over k_0^2-k_1^2}+{l_0 l_1^2\over l_0^2-l_1^2}  \right).\]

To analyse this we define $k=k_1/k_0$ with $|k|<1$ corresponding to $a>0$ at all points and similarly for $l=l_1/l_0$ and fix $k_0,l_0>0$ as the average values of $a,b$ so that we can focus on fluctuations about these as controlled by $k,l$. In this case the action becomes
\begin{equation}\label{EHkl} \int S=8\left({k_0 k^2\over 1-k^2}+ {l_0l^2\over 1-l^2}\right)=8k_0 (k^2+k^4+k^6\cdots)+8 l_0(l^2+l^4+l^6+\cdots). \end{equation}
This has a `bathtub' shape with coupling constants $k_0,l_0$ and a minimum at $k=l=0$, which makes sense as a measure of the energy of the gravitational field. The $k,l$ are not momentum variables but the relative amplitude of the unique allowed non-zero momentum in each direction. 

In the Minkowski version, we require say $a<0, b>0$ everywhere. We suppose $k_0<0,l_0>0$ as the average values  and require $|k_1|<-k_0$, $|l_1|<l_0$ to maintain the sign. We define $k,l$ as before for the relative fluctuations and regard $\tilde k_0=-k_0, l_0$ as coupling constants. Now $\mu=|ab|=-ab$ for our measure, giving
\[ \int S=8\left({\tilde k_0 k^2\over 1-k^2}- {l_0 l^2\over 1-l^2}\right)=8\tilde k_0 (k^2+k^4+k^6\cdots)-8 l_0(l^2+l^4+l^6+\cdots).\]
In either case, if we ignore higher order terms then we have $S$ quadratic in $k,l$ as for a massless free field in a universe with only one momentum in each direction. The higher terms correspond to quartic and higher derivatives in the action from this point of view.

\begin{figure}\[ 
\includegraphics[scale=.4]{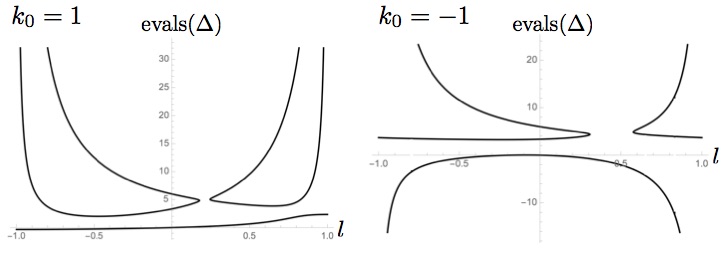}\]
\caption{Nonzero eigenvalues of the Laplacian vs $l$ at fixed $k=0.5$, $q=l_0=1$ and $k_0=\pm1$ for the two signatures.  \label{fig5}}\end{figure}

Finally, we can add matter using the Laplacian above. However, this Laplacian does depend on $q$. For example, one can check in the momentum parametrizaton that
\[ \Delta_{k_0,l_0,q; k, l}\sim \Delta_{l_0,k_0,-q; l,-k}\]
in the sense of the same eigenvalues. In other words, the theory with $a,b$ swapped is the same but has the negated value of $q$. These eigenvalues are mostly real when $q$ is real, leading to  $q=\pm1$ as the natural choices. We plot the three non-zero eigenvalues in  Figure~\ref{fig5} for $q=1$ and the two signatures, at a typical value $k=0.5$. The cross-section passes a narrow region in the $k,l$ plane where two of the eigenvalues become complex but otherwise they are positive. The remaining mainly small  eigenvalue is  zero at $l=0$ and $q=1$ or $k=0$ and $q=-1$ among possibly other zeros. 

In principle, one can proceed to consider `functional integrals' over any of our parametrizations of the metric field. Thus for quantum gravity we can consider integrals of the form
\[  \int_{-1}^1\int_{-1}^1\extd k\, \extd l\, e^{\imath \int S}\]
(this converges when we use the $\imath$ in the action, otherwise we have to renormalise due to divergence at the endpoints), and similar integrations against functions of $k,l$ to extract expectation values of operators. If we add matter to the action via the Laplacian then we will have a $q$-dependence as discussed. We should also in the full theory integrate over the $k_0,l_0$ rather than treating them as constants as we have above.

\section{Conclusions}

Sections~\ref{sec1} and~\ref{sec2} were philosophical in nature as a brief introduction to a principle of `representation-theoretic self-duality' as an `axiom of physics'\cite{Ma:pri,Ma:spbook,Ma:qg1} that has motivated many of my works. We saw how this at an abstract level was one route to the discovery of the `centre' of a monoidal category, while as `quantum Born reciprocity' it led to the discovery of an early class of quantum groups. We also explained how the big picture leads one to the {\em quantum spacetime hypothesis}. 

Sections~\ref{sec3} and~\ref{sec4}  were a brief outline of a formulation of such quantum spacetimes with curvature, using a bimodule approach  developed mostly with Edwin Beggs\cite{BegMa:sem,BegMa:cqg,BegMa:prg,BegMa:spe}. Section~\ref{sec5} then proceeded with a new application to a discrete model, namely quantum Riemannian geometry on a square. Unlike lattice approximations used in physics, we do not consider the model as an approximation but rather as an exact finite geometry\cite{Ma:gra}. We found a 1-parameter family of quantum Levi-Civita connections for every bidirectional metric   and an Einstein-Hilbert action  as a measure of the energy in the gravitational field and independent of the connection parameter. 

We also  found that the `generalised braiding' $\sigma$\cite{DubMic,Mou} emerging in our case from nothing other than the axioms of quantum Riemannian geometry applied to a square graph has a strong resemblance to the 8-vertex solutions\cite{Bax} of the Yang-Baxter equations in the theory of quantum integrable systems. Our $\sigma$ does not obey the braid relations other than the trivial case (this is usually tied to flatness of the connection\cite{Ma:gra}) but has a similar flavour. 
 
Note that while I have  focussed on my own path, reflected also in the bibliography,  there are by now many other works on quantum spacetime which I have not had room to cover.




\end{document}